\documentclass{kluwer}
\begin{document}
\begin{article}
\begin{opening}
\title{On the nature of the variable gamma-ray sources at 
low galactic latitudes}            

\author{Valent\'{\i} \surname{Bosch-Ramon}\email{vbosch@am.ub.es}}
\institute{Universitat de Barcelona}
\author{Gustavo E. \surname{Romero}}
\institute{Instituto Argentino de Radioastronom\'{\i}a\\FCAGLP, Universidad Nacional de La Plata}
\author{Josep M. \surname{Paredes}}
\institute{Universitat de Barcelona}


\runningtitle{Nature of the variable gamma-ray sources in the Galaxy}
\runningauthor{Bosch-Ramon et al.}

\begin{ao}
Valent\'{\i} Bosch-Ramon\\
Departament d'Astronomia i Meteorologia\\
Facultat de F\'{\i}sica. Universitat de Barcelona \\
Av. Diagonal 647 \\
08028 Barcelona\\
Spain
\end{ao}

\begin{abstract} 
Population studies of EGRET gamma-ray sources indicate that there is a distinctive
population of bright sources at low galactic latitudes. The sources have a distribution
consistent with that of young galactic objects, with a concentration toward the inner
spiral arms. There is a subgroup that displays strong variability 
with timescales from days to months. Following an earlier suggestion by Kaufman
Bernad\'o et al. (2002), we explore the possibility that these sources could be
high-mass microquasars. Detailed models for the gamma-ray emission that include inverse
Compton interactions of electrons in the relativistic jets and photons from all 
local fields (stellar UV photons, synchrotron photons, soft X-ray photons  from the
accretion disk, and hard X-ray photons from a corona)  are presented. We conclude that
microquasars are  excellent candidates for the parent population of the subgroup of
variable  low-latitude EGRET sources. 

\end{abstract}

\keywords{Microquasars, X-ray binaries, jets, gamma-rays}



\end{opening}

\section{Introduction}

Population studies of EGRET gamma-ray sources indicate that there is a distinctive
population of bright sources at low galactic latitudes (Gehrels et~al. 2000; Grenier 
2001, 2004; Romero 2001). They are  well-correlated with star forming regions and HII
regions, which is indicative  of an association with young stellar objects (Romero
et~al. 1999,  Romero 2001). Log $N$ -- $\log S$ studies suggest that they are  more
abundant toward the inner spiral arms (Gehrels et~al. 2000, Bhattacharya et~al. 2003).
These are bright sources (isotropic  luminosities in the range $10^{34-36}$ erg
s$^{-1}$), with an average photon  spectral index  $\left\langle\Gamma \right\rangle=2.18$ (Gehrels et~al. 2000), 
where $F(E)\propto
E^{-\Gamma}$. These  sources, whose number is $\sim 45\pm 9$, form the Gamma-Ray
Population I (GRP I, Romero et~al. 2004). Among GRP I sources there is a subgroup that
displays significant variability on  timescales of weeks to months (Torres et~al. 2001,
Nolan et~al. 2003).  Recently, Kaufman Bernad\'o et~al. (2002) and Romero et~al.
(2004)  have suggested that this subgroup of GRP I sources might be high-mass
microquasars  (i.e. microquasars formed by a compact object and an early-type stellar
companion),  where the gamma-ray emission arises from interactions between relativistic
particles  in the jet and external photon fields, most notably the stellar UV emission.

In the  present paper we explore in more detail this hypothesis, presenting more
realistic  models for the gamma-ray emission. In particular, we will include effects of
the  interaction of the microquasar jet with the X-ray fields produced by the
accretion  disk and the hot corona that is thought to surround the compact object. We
will  also include synchrotron self-Compton emission, Klein-Nishina effects, and the 
back-reaction of the different losses in the particle spectrum of the jet. We  will
calculate, for some representative sets of the parameters that characterize high-mass
microquasars, the spectral energy distribution (SED) from radio wavelengths up to GeV
gamma-rays. 

\section{GRP I sources}

GRP I sources concentrate along the galactic plane and present a good spatial correlation 
with young stellar objects (Romero et~al. 1999). The variability analysis of 
these sources by Torres et~al. (2001) clearly shows evidence for the existence 
of a subgroup with variable emission on timescales from weeks to months. This is 
corroborated by the recent results presented by Nolan et~al. (2003), which 
are based on a maximum likelihood re-analysis of the EGRET data. These authors identify 17 
variable sources within $6^{\circ}$ from the galactic plane. These sources are clumped 
within $55^{\circ}$ of the galactic center.

A $\log N - \log S$ analysis for all GRP I sources yields a distribution that is 
consistent with a power-law with index $\left\langle\beta \right\rangle=3.1\pm0.4$ (Bhattacharya et~al. 
2003). 
This is far steeper than what is expected for a population uniformly 
distributed along the galactic disk. For instance, for pulsars detected at 400 MHz the 
slope results $\left\langle\beta \right\rangle\sim1.05$ (Bhattacharya et~al. 
2003). The unidentified gamma-ray sources, on the contrary, 
seem to be concentrated mainly in the inner spiral arms. In order to find possible 
further evidence for different populations among GRP I sources, we have implemented a 
$\log N - \log S$ analysis of both variable and non-variable low-latitude sources.

\begin{figure}
\vspace{7cm}
\includegraphics{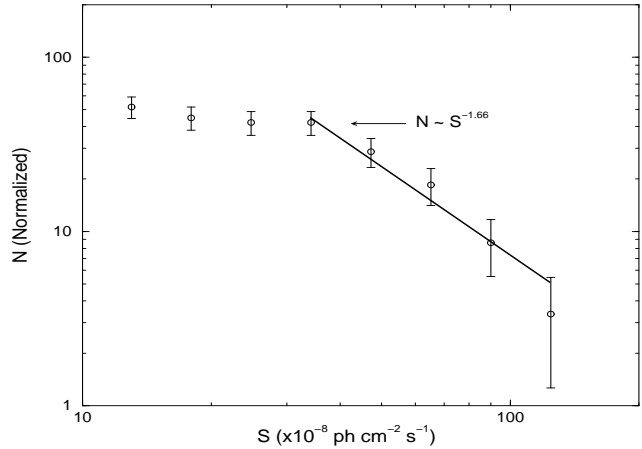}
\caption{Log $N - \log S$ plot for variable gamma-ray sources within $|b|<6^{\circ}$.}
\label{Fig1}
\end{figure}

\begin{figure}
\vspace{7cm}
\includegraphics{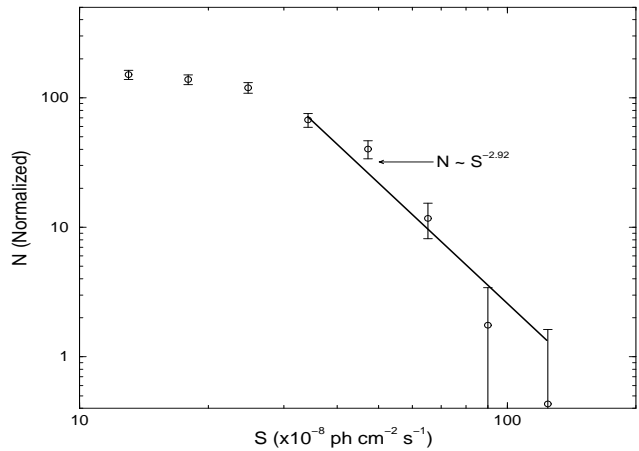}
\caption{Log $N - \log S$ plot for non-variable gamma-ray sources within $|b|<6^{\circ}$.}
\label{Fig2}
\end{figure}

First we have considered the 17 variable sources listed by Nolan et~al. (2003). 
In order to take into account systematic effects introduced by different exposure and 
background resulting in non-uniform detectability, we have adopted the procedure 
described by Reimer (2001). The obtained $\log N - \log S$ plot is shown 
in Fig.~\ref{Fig1}. The normalized distribution can be fitted by a power-law 
$N(S)\propto S^{-\beta}$, with  $\left\langle\beta \right\rangle=1.66\pm0.31$, significantly flatter than for 
the entire sample. If we now consider those sources that classified as non-variable 
or dubious cases, we get the $\log N - \log S$ plot shown in Fig.~\ref{Fig2}. In 
this case the distribution can be fitted by a power-law with index $
\left\langle\beta \right\rangle=2.92\pm0.36$. Quoted errors for $\beta$ are 1 sigma errors.      

The average spectral index is also different for both samples: in the case of the variable 
sources we have $\left\langle \Gamma \right\rangle=2.04\pm 0.1$, whereas for the remaining 
sources is $\left\langle \Gamma \right\rangle=2.16\pm 0.2$. However, the averaged errors are
still large, due to the low number of sources, and do not allow yet to distinguish both
populations properly. In any case, $\log N - \log S$ analysis suggest that there 
are two different groups of sources, one formed by steady sources concentrated toward 
the inner spiral arms, and a second group with variable sources and a wider distribution 
along the galactic plane, although not as wide as that of radio pulsars (Bhattacharya 
et~al. 2003).

High-mass microquasars appear to be good candidates for compact and variable sources in the 
galactic plane. Since they can have very large proper motions (e.g. Rib\'o et~al. 
2002), their distribution along the plane should be broader than that presented 
by supernova remnants and molecular clouds, which can be traced by star-forming regions 
and OB associations. Their spread, however, is limited by the lifespan of the companion 
massive star, and hence it is not as extended as that of radio pulsars. In the next 
section we will discuss the potential of microquasars as gamma-ray sources.

\section{Gamma-ray emission from microquasars}

In the present paper we will consider a relativistic inhomogeneous jet flow moving along the $z$
axis (assumed to be perpendicular to the disk) with a bulk Lorentz factor $\Gamma_{\rm jet}$. We
shall allow the jet to expand laterally, in such a way that  the radius $R$ at a distance $z$
from the compact object will be given by  $R(z)=\xi z^{\varepsilon}$, with $\varepsilon\leq 1$
and $z_0\leq z\leq z_{\rm max}$. $\xi$ is related to the opening angle of the jet. For $\epsilon=1$
we have a conical jet. The electron energy distribution  ($N(z,\gamma_{\rm e})$, a power-law of
index $p$), the maximum electron Lorentz factor ($\gamma_{\rm e max}(z)$) and the magnetic field
($B(z)$) will be parametrized like in Ghisellini et~al. (1985). 

The synchrotron radiation density within the jet can be estimated using the local approximation to
the synchrotron radiation field (e.g. Ghisellini et~al. 1985). We have used the formulae presented
in Pacholczyk (1970), adapted to an  expanding jet. The total radiation field to which the leptons
are exposed in the jet will have also  contribution from external sources. These contributions can
be modeled as two black body  components, one peaked at UV energies (the companion star field) and
other at energies  $kT\sim 1$ keV (the inner accretion disk field), plus a power-law with an
exponential  cutoff at $kT\sim 150$ keV (the corona). With the exception of the disk,  these
contributions  are assumed to be isotropic (Romero et~al. 2002, Georganopoulos  et~al.  2002).

Once $N(z,\gamma_{\rm e})$ and the total energy density in seed photons  ($U_{\rm
tot}(\epsilon_0,z)$) have been determined, the inverse Compton (IC) interaction between them can be
studied calculating the IC spectral energy distribution per energy unit. The cross
section ($\sigma(x,\epsilon_0,\gamma_{\rm e})$) for both the Thomson and the Klein-Nishina 
regimes has been taken from
Blumenthal \& Gould (1970). The spectral energy distribution for the optically thin
case in the jet's reference frame is:
\begin{eqnarray}
\epsilon L_{\epsilon}&=&\epsilon
\int^{z_{\rm max}}_{z_{\rm min}}
\int^{\epsilon_{\rm 0 max}(z)}_{\epsilon_{\rm 0 min}(z)}
\int^{\gamma_{\rm emax}(z)}_{\gamma_{\rm emin}(z)}
\Sigma(z)U_{\rm tot}(\epsilon_0,z) \nonumber \\ &&
\times  N(\gamma_{\rm e},z)\frac{d\sigma(x,\epsilon_0,\gamma_{\rm e})}{d\epsilon}
\frac{\epsilon}{\epsilon_0}d\gamma 
d\epsilon_0 
dz,
\label{eq:Lic}
\end{eqnarray}
where $\epsilon_0$ and $\epsilon$ are the energies of the incoming and the outgoing
photons, respectively. The parameter $x$ is actually a function which depends on both of the former
quantities and on the electron Lorentz factor. $\Sigma(z)$ is the surface of a perpendicular jet slice located at $z$.
Notice that the external fields contributing to $U_{\rm tot}(\epsilon_0,z)$ 
should be transformed to the co-moving frame. Detailed expressions for such 
transformations are given by Dermer \& Schlickeiser (2002). 

In the observer's reference frame we have:
\begin{eqnarray}
\epsilon' L'_{\epsilon'}&=& D^{2+p} \epsilon' L_{\epsilon'}.
\label{eq:LicSRobs}
\end{eqnarray}
The integration is performed in the co-moving system and then the result is transformed
to the  observer's frame, hence the factor $D^{2+p}$, which is the Doppler boosting for
a continuous  jet. The energy of the scattered photons in the jet's reference frame
($\epsilon$)  is boosted to $\epsilon'=D\epsilon$. The Doppler factor $D$ for the 
approaching jet is given by  
\begin{equation}
D=\frac{1}{\Gamma_{\rm jet}(1-\beta\cos\theta)},
\end{equation}
where $\beta$ is the velocity of the jet in speed of light units and $\theta$ is the
angle between the jet and the line of sight. 

In the case of the IC interactions with disk photons, a factor $(1-\cos
\theta)^{(p+1)/2}$ must  be introduced in Eq.(~\ref{eq:Lic}) in order to take into
account the fact that the photons come from {\it behind} the jet (Dermer et~al.
1992).  

In order to make any calculation of the IC emission of a given microquasar, we have to specify 
first the jet power. In this work we 
shall adopt the disk/jet coupling hypothesis formulated by Falcke \& Biermann  
(1995, 1999), i.e. the jet power is proportional 
to the accretion rate:
\begin{equation}
L_{\rm jet}=qL_{\rm ac}.
\end{equation}
Here, $L_{\rm ac}=\dot{M}c^2$ is the accretion power onto the compact object and $q$ 
is a number $<1$. For a leptonic jet, as in our case, $q\ll 1$. 
The jet power determines the energy content of the electron energy distribution for a leptonic
jet (see, e.g., Georganopoulos et~al. 2002).

Although the model outlined above concerns primarily to 
the high energy emission range since 
it focuses on the IC radiation from the inner jet, the synchrotron formulae can be used to 
make predictions also at radio wavelengths. With this aim, we have developed an outer jet 
model appropriate for regions farther from the compact object. This outer jet model will 
fulfill most of the conditions of the former one, but it will require a different geometry 
because of the expected lower energy loss rate of the particles. 

For further details about our model, see Bosch-Ramon et~al. (2004).

\section{Specific models} \label{exploring}

In order to explore a plethora of situations, we have adopted a set of assumptions, which are
summarized in Tables~\ref{paramvalues}. In Table~\ref{tideal} we list those
parameters that were allowed to change along with the values adopted for three representative
cases: a `mild' microquasar, a microblazar, and a system similar to LS 5039 and LS~I~+61~303\footnote{LS~5039 and
LSI~+61$^{\circ}$303 present significant differences though, regarding gamma-rays, they are pretty
similar. Also, at lower energies, both systems can be treated in a similar way at this stage.}.  

\begin{table*}
\begin{flushleft}
\caption[]{Main fixed parameters in the models.}
\begin{tabular}{l c c c c c}
\noalign{\smallskip} \hline \noalign{\smallskip} Parameter (symbol) & Value 
\cr \noalign{\smallskip} \hline \noalign{\smallskip} 
Black hole mass ($M_{\rm bh}$) & $10 M_{\odot}$  
\cr Gravitational radius ($R_{\rm g}$) & $1.48\times10^6$~cm  
\cr Accretion luminosity ($L_{\rm ac}$) & $10^{-8}~M_{\odot}$c$^2$~year$^{-1}$
\cr Stellar radius ($R_{*}$) & $15~R_{\odot}$  
\cr Stellar bolometric luminosity ($L_{*}$) & 5$\times10^{38}$~erg~s$^{-1}$
\cr Distance from jet's apex to the compact object ($z_0$) & $50~R_{\rm g}$ 
\cr Initial jet radius ($R_0$) & 0.1$z_0$
\cr Orbital radius ($R_{\rm orb}$) & $45~R_{\odot}$
\cr Peak energy of the disk ($kT_{\rm disk}$) & 1~keV
\cr Peak energy of the corona ($kT_{\rm cor}$) & 150~keV
\cr Peak energy of the star ($kT_{\rm star}$) & 10~eV
\cr Expansion coefficient of the inner jet ($\varepsilon$) & 1
\cr Electron power-law index ($p$) & 2
\cr Photon index for the corona ($\Gamma_{\rm cor}$) & 1.6
\cr \noalign{\smallskip} \hline
\end{tabular}
\label{paramvalues}
\end{flushleft}
\end{table*}

\begin{table}
\begin{flushleft}
\caption[]{Changing parameters in the different cases considered in the text.}
\begin{tabular}{l c c c c c}
\noalign{\smallskip} \hline \noalign{\smallskip} Parameter (symbol) 
& Value \\
 & mild/extreme/realistic
\cr \noalign{\smallskip} \hline \noalign{\smallskip} 
Corona luminosity ($L_{\rm cor}$) & $10^{35}$, 
$10^{35}$, $3\times 10^{32}$~erg~s$^{-1}$
\cr Total disk luminosity ($L_{\rm disk}$) & $10^{37}$, $10^{37}$, 5$\times10^{32}$~erg~s$^{-1}$
\cr Inner jet magnetic field ($B(z_0)$) & 10, 10, 200~G
\cr Maximum electron Lorentz factor ($\gamma_{\rm emax}(z_0)$) & 10$^4$, 10$^6$, 10$^4$
\cr Bulk Lorentz factor of the jet ($\Gamma_{\rm jet}$) & 2.5, 10 , 1.1
\cr Viewing angle to jet's axis $\theta$ & $10^{\circ}$, $1^{\circ}$, $10^{\circ}$
\cr Disk/jet coupling constant ($q$) & $10^{-4}$, $10^{-4}$, $10^{-3}$
\cr \noalign{\smallskip} \hline
\end{tabular}
\label{tideal}
\end{flushleft}
\end{table}

\begin{table*}
\begin{flushleft}
\caption[]{Outer jet parameters (different cases considered in the text).}
\begin{tabular}{l c c c c c}
\noalign{\smallskip} \hline \noalign{\smallskip} Parameter (symbol) & Value \\
 & mild/extreme/realistic
\cr \noalign{\smallskip} \hline \noalign{\smallskip} 
Expansion coefficient of the outer jet ($\varepsilon$) & 0.2, 0.3, 0.2
\cr Initial point of the outer jet ($z_0$) & $500~R_{\rm g}$,
$5000~R_{\rm g}$ , $500~R_{\rm g}$
\cr Initial radius of the outer jet ($R_0$) &  $50~R_{\rm g}$, $500~R_{\rm g}$, $50~R_{\rm g}$
\cr (Computed) maximum length of the jet (few~GHz) &  $\sim 10^{16}$, $\sim 10^{17}$, 
$\sim 10^{15}$~cm
\cr Maximum electron lorentz factor of the outer jet ($\gamma_{\rm emax}$) & $10^3$, 
$10^4$, $10^3$
\cr Outer jet magnetic field ($B(z_0)$) & 1, 0.1, 20~G
\cr Electron power-law index ($p$) & 2
\cr \noalign{\smallskip} \hline
\end{tabular}
\label{radio}
\end{flushleft}
\end{table*}

\begin{figure}
\vspace{7cm}
\includegraphics{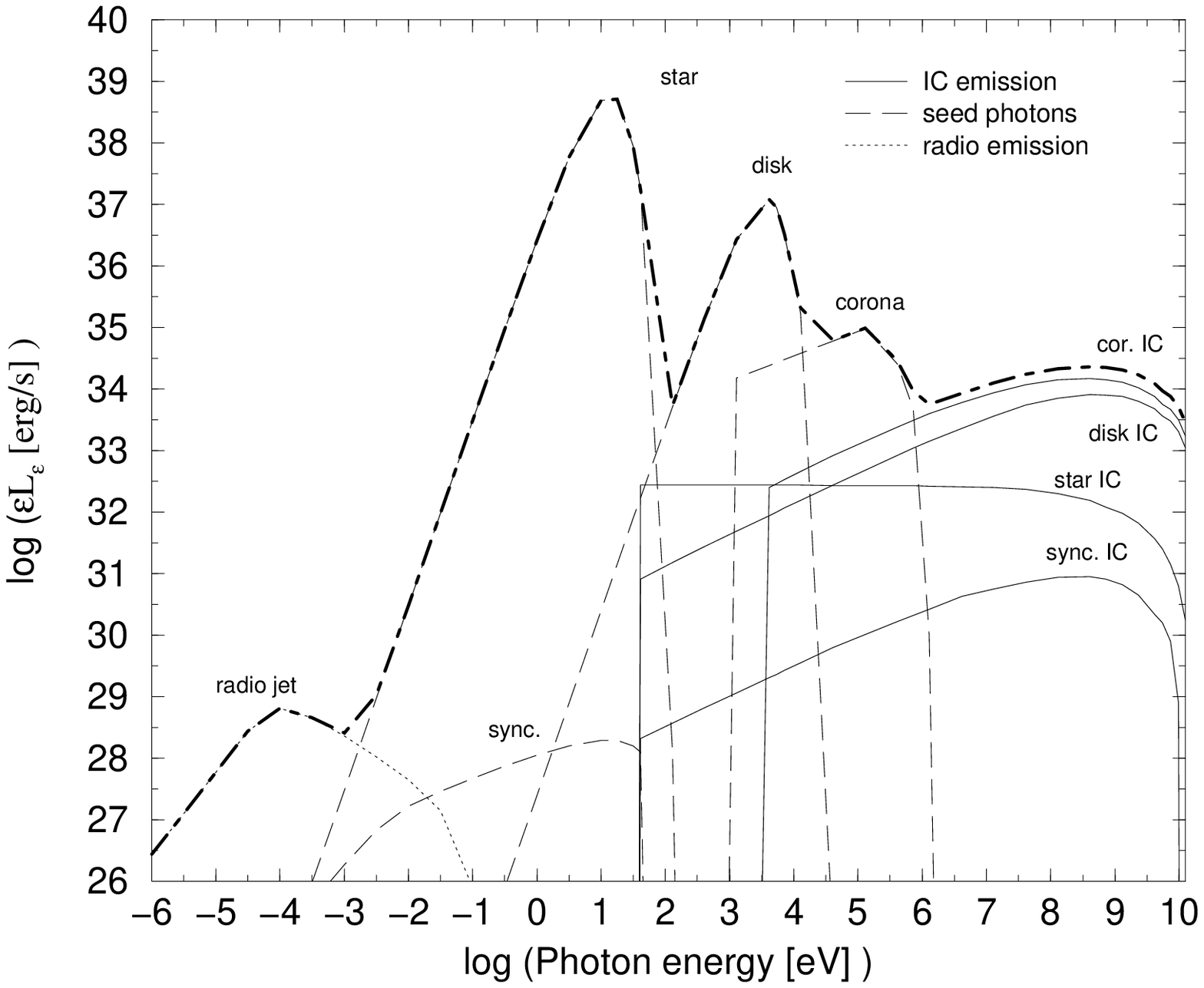}
\caption{Complete SED for a model with $q=10^{-4}$, $B$=10~G, $L_{\rm disk}=10^{37}$~erg~s$^{-1}$,
$L_{\rm cor}=10^{35}$~erg~s$^{-1}$, $\gamma_{\rm emax}(z_0)=10^4$, $\Gamma_{\rm jet}$=2.5, 
and a viewing angle of $10^{\circ}$. The synchrotron seed photons come from the inner jet.} 
\label{ideal}
\end{figure}

\begin{figure}
\vspace{7cm}
\includegraphics{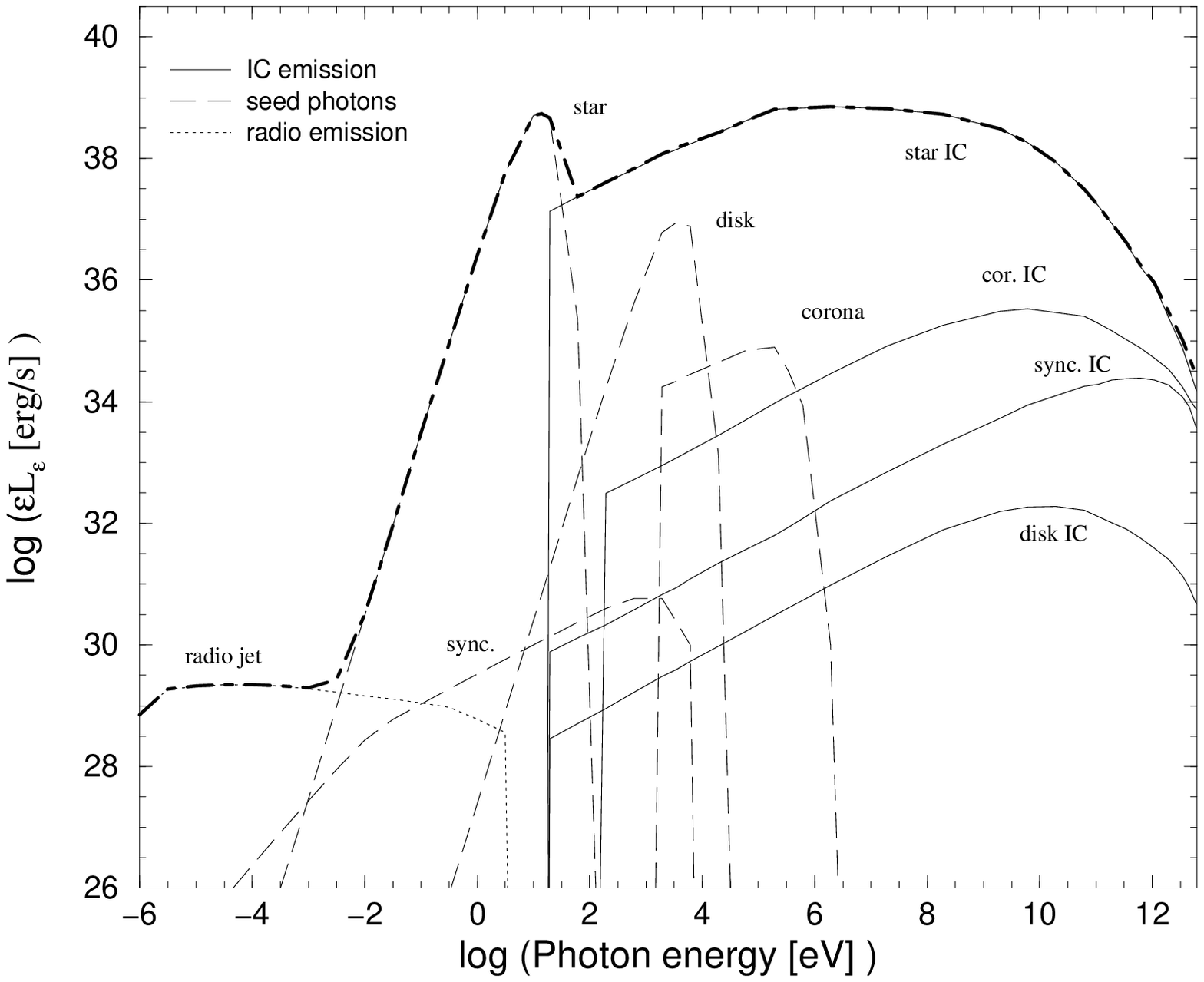}
\caption{SED for a model with $q=10^{-4}$, $B$=10~G, $L_{\rm disk}=10^{37}$~erg~s$^{-1}$, 
$L_{\rm cor}=10^{35}$~erg~s$^{-1}$, $\gamma_{\rm emax}(z_0)=10^6$, $\Gamma_{\rm jet}$=10, 
and a viewing angle of $1^{\circ}$. The synchrotron seed photons come from the inner jet.} 
\label{extreme}
\end{figure}

\begin{figure}
\vspace{7cm}
\includegraphics{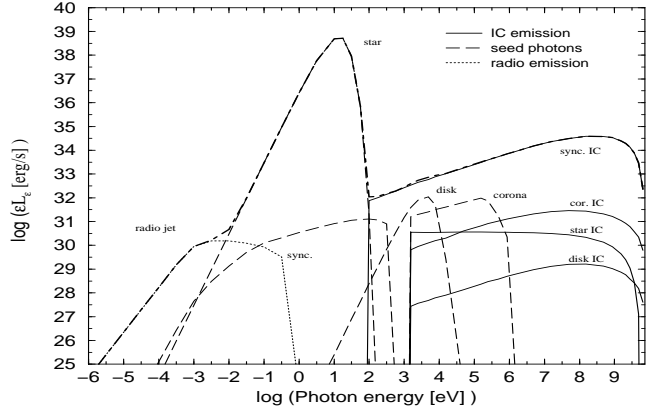}
\caption{SED for a model with $q=10^{-3}$, $B$=200~G, $L_{\rm
cor}$ and $L_{\rm disk}=3\times 10^{32}$~erg~s$^{-1}$, $\gamma_{\rm emax}(z_0)=10^4$, 
$\Gamma_{\rm jet}$=1.1, and 
a viewing angle of $10^{\circ}$. The synchrotron seed photons come from the inner jet.} 
\label{real}
\end{figure}

In Fig.~\ref{ideal}, we show the whole SED of the `mild' case. We plot here the spectra for all contributions  to the emission, i.e. the
outer jet, the seed photon sources, and the IC components  (for outer jet parameters, see
Table~\ref{radio}). In Fig.~\ref{extreme} we show a more  extreme case, for a source with a high
bulk Lorentz factor $\Gamma_{\rm jet}$=10, a high-energy  cutoff for electrons of $\gamma_{\rm
emax}(z_0)=10^6$, and a small viewing angle $\theta=1^{\circ}$ (see Table~\ref{tideal}). Finally,
we plot a SED similar (realistic case in Tables~\ref{tideal} and \ref{radio}) to the one presented by two microquasars that are considered as the best 
candidates to counterparts of unidentified EGRET sources, LS~5039 and LSI~+61$^{\circ}$303
(3EG~1824+1514, Paredes et~al. 2000, and 3EG~0241+6103, Kniffen et~al. 1997, respectively -- see
Fig.~\ref{real}); the selected set of parameter is consistent with those adopted in Bosch-Ramon \& Paredes (2004a, b).
\newpage

\section{Discussion} \label{disc}

In a previous work (Bosch-Ramon et~al. 2004), it was shown that  for several models with disk/jet
coupling constant $q=10^{-3}$ we can get the expected luminosities in the observer's  frame
inferred for GRP I sources with the right photon index at energies $\sim1$~GeV,  i.e. $\Gamma\sim
2$. When the magnetic field is strong enough ($B(z_0)\sim 100$ G),  SSC emission alone can account
for luminosities $\sim 10^{35}$ erg s$^{-1}$ at 1 GeV.  Models with bulk Lorentz factors
$\Gamma_{\rm jet}=1.5$ and $5$ do not produce  dramatically different results for a jet with a
viewing angle of $\sim 10^{\circ}$  except for the case of the scattered stellar photons. In case
of lower magnetic fields,  the IC scattering upon external  fields clearly dominates. It is
interesting to notice that even models with light jets ($q=10^{-5}$) can produce significant
gamma-ray sources ($\sim 10^{34}$ erg s$^{-1}$ at 100~MeV) when a strong corona is present. In
these models, the spectrum tends to be a bit harder than in the case of stellar photons, with our
current set of assumptions (value of $\Gamma_{\rm cor}$, etc).

In the present study of concrete cases, we compare Figs.~\ref{ideal},
~\ref{extreme}~and~\ref{real}. From these figures, it is possible to see the
differences between `mild' microquasars, with $\Gamma_{\rm jet}=2.5$, a viewing
angle of $\theta=10^{\circ}$ and $q=10^{-4}$, and more `extreme' cases with
$\Gamma_{\rm jet}=10$ and $\theta=1^{\circ}$. Since we have assumed in both
cases a magnetic field of $B(z_0)=10$~G, external IC scattering always
dominates over SSC. In the `mild' microquasar the emission at MeV-GeV energies
is mainly due to upscattering of coronal photons whereas, in the extreme case,
the IC emission in the stellar field exceeds by far the other contributions.
For the case similar to LS~5039/LSI~+61$^{\circ}$303, we have taken the
magnetic field to be high enough to be the main contribution at high energies
to reproduce the observed features of the spectra; mainly: higher luminosities
at gamma-rays than at X-rays and no strong disk/corona hints at X-rays.  In the
three cases the radio counterpart is rather weak, with a $L_{\rm
radio}/L_{\gamma}\ll 1$. The hard X-ray counterpart, on the contrary, can be
important, even beyond the cutoff for the corona (at $\sim 150$~keV). Hence,
INTEGRAL observations can be useful to unmask candidates that are obscured at
optical wavelength (see Combi et~al. 2004 for a recent study in this
direction). 

An interesting feature of the extreme case is that the gamma-ray emission with a hard spectrum
$\Gamma\sim 2$ extends up to $\sim$100~GeV. Above $10^{11}$~eV the spectrum becomes softer due
to the Klein-Nishina effect. This type of sources should be detectable with modern Imaging
Cherenkov Telescopes like HESS and MAGIC. Even systems with low-mass companions and weak coronas
might be detectable since the SSC component has luminosities of $\sim 10^{35}$~erg~s$^{-1}$. The
soft spectrum can be used to differentiate this leptonic model from other hadronic alternatives
which predict a harder spectrum at very high energies (e.g. Romero et al.
2003).                   

\section{Conclusions}

We have shown that the variable gamma-ray sources found on the galactic plane have some common
features that make reasonable to consider them as a distinctive group of GRP I sources. We have
suggested that these sources might be microquasars with high-mass stellar companions and we have
elaborated some detailed models to explain the gamma-ray production in this type of objects. In
particular, we have considered inhomogeneous jet models where gamma-rays are the result of inverse
Compton interactions of leptons in the inner jet with locally produced synchrotron photons as well
as external photon fields. We have calculated the emission resulting from the upscattering of
disk, coronal, and stellar photons, incorporating a full Klein-Nishina calculation and the effect
of losses in the particle spectrum. We have shown that different spectral energy distributions at
high energies can be obtained from different and reasonable combinations of the physical
parameters like magnetic field, jet power, coronal and disk luminosities, etc. The model can
compute with a reasonable set of parameters a realistic SED, reproducing the ones associated to
known microquasars. It seems to be clear that the microquasar phenomenon can be naturally extended up to
the highest energies and that we can expect these objects to manifest themselves as a distinctive
group of gamma-ray sources that might be detectable with satellite-borne instruments like those
to  be carried by AGILE and GLAST and even by ground-based Cherenkov telescopes like HESS and
MAGIC.      

Microquasars are not, of course, the only kind of galactic object that might display
variable gamma-ray emission. Other alternatives include early-type binaries (Benaglia \&
Romero 2003), accreting neutron stars (Romero et~al. 2001), pulsar wind nebulae (Roberts
et~al. 2002) and exotic objects (Punsly et~al. 2000). However, microquasars are perhaps the
most attractive candidates to explain a significant fraction of the variable GRP I sources
because of the well-established presence of relativistic jets in these objects, along with several external photon fields.

\acknowledgements
J.M.P. and V.B-R. acknowledge partial support by DGI of the Ministerio de
Ciencia y Tecnolog\'{\i}a (Spain) under grant AYA-2001-3092, as well as
additional support from the European Regional Development Fund (ERDF/FEDER).
During this work, V.B-R. has been supported by the DGI of the Ministerio de
Ciencia y Tecnolog\'{\i}a (Spain) under the fellowship FP-2001-2699. G.E.R. is supported by Fundaci\'on Antorchas and the Argentine Agencies CONICET and ANPCyT. 
This research has benefited from a cooperation agreement supported by
Fundaci\'on Antorchas. We thank an anonymous referee by for constructive comments on manuscript.

\end{article}

\begin{thebibliography}{}

\bibitem[]{}
Benaglia, P., Romero, G.E. 2003, A\&A 399, 1121

\bibitem[]{} 
Bhattacharya, D., Aky\"uz, A. Miyagi, T., Samimi, J. ~\& Zych, A. 2003, A\&A 404, 163 

\bibitem[]{}
Blumenthal, G. R.~\& Gould, R. J. 1970, Rev. Mod. Phys. , 42, 237

\bibitem[]{}
Bosch-Ramon, V.~\& Paredes, J.~M.\ 2004a, A\&A, 417, 1075 

\bibitem[]{}
Bosch-Ramon, V.~\& Paredes, J.~M.\ 2004b, A\&A, 425, 1069
 
\bibitem[]{}
Bosch-Ramon V., Romero, G.~E.~\& Paredes, J.~M.\ 2005, A\&A, 429, 267

\bibitem[]{}
Combi, J.~A., Rib\'o, M., Mirabel, I.~F., \& Sugizaki, M.
2004, A\&A, 422, 1031

\bibitem[]{}
Dermer, C.~D., Schlickeiser, R., \& Mastichiadis, A.\ 1992, A\&A, 256, L27

\bibitem[]{} Dermer, 
C.~D.~\& Schlickeiser, R.\ 2002, ApJ, 575, 667 

\bibitem[]{}
Falcke, H.~\& Biermann, P.~L.\ 1995, A\&A, 293, 665

\bibitem[]{}
Falcke, H.~\& Biermann, P.~L.\ 1999, A\&A, 342, 49

\bibitem[]{} 
Gehrels, N., Macomb, D.J., Bertsch, D.L., Thompson, D. J., \& Hartman R.C. 2000, 
Nat 404, 363

\bibitem[]{}
Georganopoulos, M., Aharonian, F.~A., \& Kirk, J.~G. 2002, A\&A, 388, L25

\bibitem[]{}
Ghisellini, G., Maraschi, L., \& Treves, A. 1985, A\&A, 146, 204 

\bibitem[]{} 
Grenier, I.A. 2001, in: The Nature of Unindentified Galactic High-Energy
Gamma-Ray Sources, ed. A. Carraminana, O. Reimer, \& D. Thompson, Kluwer
Academic Publishers, Dordrecht, p.51

\bibitem[]{}
Grenier, I.A. 2004, in: Cosmic Gamma-Ray Sources, ed. K.S. Cheng \& G.E. Romero, Kluwer
Academic Publishers, Dordrecht, p.47

\bibitem[]{}
Kaufman Bernad\'o, M.~M., Romero, G.~E., \& Mirabel, I.~F. 2002, A\&A, 385,
L10

\bibitem[]{}
Kniffen, D.~A., Alberts, W.~C.~K., Bertsch, D.~L.~et al.\ 1997, ApJ, 486, 126

\bibitem[]{}
Nolan, P.L, Tompkins, W.F., Grenier, I.A., Michelson, P.F. 2003, ApJ, 597, 615 

\bibitem[]{}
Pacholczyk, A.~G., 1970, Radio Astrophysics, Freeman, San Francisco, CA

\bibitem[]{}
Paredes, J.~M., Mart{\'{\i}}, J., Rib\'o, M., \& Massi, M. 2000, Science, 288, 2340

\bibitem[]{}
Punsly, B., Romero, G.E., Torres, D. F., Combi, J. A. 2000, A\&A, 364, 552

\bibitem[]{} 
Reimer, O. 2001, in: The Nature of Unindentified Galactic High-Energy
Gamma-Ray Sources, ed. A. Carraminana, O. Reimer, \& D. Thompson, Kluwer
Academic Publishers, Dordrecht, p.17

\bibitem[]{}
Rib\'o M., Paredes J.M., Romero G.E., et~al., 2002, A\&A, 384, 954

\bibitem[]{}
Roberts, M.S.E.; Gaensler, B.M., Romani, R.W. 2002,
in: Neutron Stars in Supernova Remnants, ASP Conference Series, Vol. 271, Patrick O. 
Slane and Bryan M. Gaensler (eds.), San Francisco: ASP, 2002., p.213

\bibitem[]{}
Romero, G.~E., Benaglia, P., Torres, D.~F. 1999, A\&A, 348, 868

\bibitem[]{}
Romero, G.~E. 2001, in: The Nature of Unindentified Galactic
High-Energy Gamma-Ray Sources, ed. A. Carraminana, O. Reimer, \& D. Thompson, Kluwer
Academic Publishers, Dordrecht, 65

\bibitem[]{}
Romero, G.E., Kaufman Bernad\'o, M.M., Combi, J.A., Torres, D.F. 2001, A\&A, 376, 599

\bibitem[]{}
Romero, G.E., Kaufman Bernad\'o, M.M., \& Mirabel, I.F. 2002,
A\&A, 393, L61

\bibitem[]{}
Romero, G.~E., Torres, D.~F., Kaufman  Bernad\'o, M.~M., \& Mirabel, I.~F. 2003, A\&A, 
410, L1  

\bibitem[]{}
Romero, G.~E., Grenier, I.~A., Kaufman Bernad\'o, M.M., \& Mirabel, I.F., \& Torres, 
D.~F. 2004, ESA-SP, 552, 703

\bibitem[]{} 
Torres, D.~F., Romero, G.~E., Combi, J.~A., Benaglia, P., Andernach, H., \& Punsly, 
B.\ 2001, A\&A, 370, 468 

\end{thebibliography}
\end{document}